УДК 538.9

P.V.Gorskyi

# ON CONDITIONS OF HIGH FIGURE OF MERIT AND METHODS OF SEARCH FOR PROMISING SUPERLATTICE THERMOELECTRIC MATERIALS

Institute of Thermoelectricity of the NAS and MES Ukraine,
1 Nauky str., Chernivtsi, 58029, Ukraine

*This paper presents a rigorous calculation of the figure of merit of superlattice thermoelectric material (SL TEM) with regard to real three-dimensionality and nonparabolicity of its energy spectrum with the arbitrary level of openness of its Fermi surface (FS). In this case, the quantitative characteristic of the level of openness of FS is the ratio of the Fermi energy of ideal two-dimensional Fermi gas with a square law of dispersion at absolute zero temperature to half-width of a narrow miniband defining charge carrier motion between SL TEM layers. In so doing, the law of dispersion of charge carriers in a narrow miniband is assumed to be corresponding to the Fivaz model, i.e. cosine. In the calculation of figure of merit it is also assumed that the mean free path of charge carriers does not depend on their energy, hence on quantum numbers, and the lattice component of thermal conductivity obeys the law of Leibfried and Shlemann, i.e. is inversely proportional to temperature. It is shown that the figure of merit of SL TEM in the temperature range of 300-500K is drastically increased with increasing level of openness of FS. Therefore, even for closed or transient FS with the level of openness close to or equal to unity with optimal combination of parameters one can obtain such figure of merit of SL TEM which at extreme temperatures 300 and 500K assures generator efficiency equal to 34%, which is 85% of the Carnot cycle efficiency. However, due to the presence of lattice component of thermal conductivity the figure of merit of SL TEM is rather responsive to the distance between the layers and drastically drops with its increase. Therefore, with increasing this distance beyond the optimum by a factor of 5, the generator efficiency drops to 4%, which makes as low as 10% of the Carnot cycle efficiency.*

*Besides, for the same material, coefficient of performance of a refrigerator in the framework of its simplest model was calculated. It was established that if material band spectrum is described by the Fivaz model, the coefficient of performance is drastically increased with increasing the level of openness of FS and in case of transient FS it reaches 1.9 between the temperatures of 300 and 230K. However, as the distance between the layers of SL TEM increases 2 times, it drops to 0.4, and the level of openness of FS, necessary for the onset of cooling, is drastically increased. Nevertheless, due to proper selection of charge carrier concentration one can increase considerably both generator efficiency and coefficient of performance of a refrigerator even for SL TEM with a large lattice constant.*

*Based on the obtained criteria, four methods of search for promising SL TEM with the use of quantizing magnetic fields are proposed. The first method lies in the investigation of de Haas-Shubnikov oscillations at helium temperatures and selection on this basis of materials with the level of openness of FS close to or equal to unity. The second method lies in selection of such materials at helium temperatures in close to ultra-quantum fields according to the presence of a well-expressed area of negative magnetoresistance. The third method consists in selection of such materials at helium temperatures in close to ultra-quantum magnetic fields, according to the degree of roundness of power factor peak. The fourth method consists in selection of such materials at higher temperatures according to well-expressed negative longitudinal magnetoresistance.*

**Introduction**

Thermoelectric figure of merit of material is one of the key parameters defining the advisability of its application for creation of thermoelectric energy converters. The figure of merit of thermoelectric materials is mainly improved in two ways. The first way is to improve conventional materials, for instance, solid solutions of *Bi(Sb)-Te(Se)* system due to their optimization for concentration of doping impurities [1] or due to the use of classical and (or) quantum dimensional effects in passing from single crystals to thin films, wires or powder based materials [2-5]. The second way lies in devising (creation) of "nonconventional" materials, where the specific features of band spectrum of charge carriers give grounds for expecting high values of thermoelectric figure of merit. As such materials, the so-called "superlattice" materials with pronounced layered structure and carrier band spectrum anisotropy [6-8], for instance, materials on the basis of Al-Ga-As or Si-Ge system, widely used in electronics and optics [9], have been mush investigated. The expectations for high thermoelectric figure of merit of such materials are due to the fact that charge carriers in these materials are strongly localized in the layers, by virtue of which charge carrier band spectrum is "nearly" two-dimensional. And this, at least due to a drastic increase in the density of states and, hence, thermopower, should have resulted in considerable improvement of thermoelectric figure of merit of SL TEM. However, abundant evidence [10-15] shows that in reality the thermoelectric figure of merit of these materials does not exceed that of conventional materials or even is much lower. Relatively high *ZT* values of the order of 2-4 at 300K in the existing or investigated SL TEM are rare in occurrence.

In view of the foregoing, the purpose of the present paper is a sound estimate of the thermoelectric figure of merit of SL TEM, analysis of factors contributing to its improvement or reduction, and development of methods of search for promising SL TEM with the use of quantizing magnetic fields.

**General formula for the figure of merit of SL TEM and its analysis**

In the framework of the Fivaz model [16] the band spectrum of charge carriers in SL TEM is of the form:

$$\varepsilon(\vec{k}) = \frac{\hbar^2 k_\parallel^2}{2m^*} + \Delta(1 - \cos ak_\perp), \qquad (1)$$

where $k_\parallel, k_\perp$ – quasi pulse components in layers plane and normal to it, respectively, $m^*$ – effective mass of charge carriers in the direction normal to layers, $\Delta$ – miniband half-width in the direction normal to layers, $a$ – the distance between translation equivalent layers.

In our calculations we will assume temperature gradient and electric field to be parallel to each other and to layers plane. With this configuration, on the assumption of energy independence of charge carrier means free path, the dimensional thermoelectric figure of merit of SL TEM in the impurity region at temperature $T$ with regard to the contribution of lattice thermal conductivity is determined as follows:

$$ZT = 8\pi^3 \frac{A_0}{B_0 + c_l \left(2k_B^2 \zeta_{02D}^2 l\right)^{-1} ah^2 \sqrt{m^{*-1} \zeta_{02D}}}. \qquad (2)$$

Dimensionless functions $A_0$ and $B_0$ are given by:

$$A_0 = \left\{ \int\limits_0^\infty \int\limits_0^\pi \frac{y\left[y + K^{-1}(1-\cos x) - \gamma^*\right] \exp\left\{\left[y + K^{-1}(1-\cos x) - \gamma^*\right]/t_{2D}\right\}}{\left\{\exp\left\{\left[y + K^{-1}(1-\cos x) - \gamma^*\right]/t_{2D}\right\} + 1\right\}^2 \sqrt{2y + 4\pi K^{-2} n_0 a^3 \sin^2 x}} dxdy \right\}^2 \times$$
$$\left\{ \int\limits_0^\infty \int\limits_0^\pi \frac{y \exp\left\{\left[y + K^{-1}(1-\cos x) - \gamma^*\right]/t_{2D}\right\}}{\left\{\exp\left\{\left[y + K^{-1}(1-\cos x) - \gamma^*\right]/t_{2D}\right\} + 1\right\}^2 \sqrt{2y + 4\pi K^{-2} n_0 a^3 \sin^2 x}} dxdy \right\}^{-1}. \quad (3)$$

$$B_0 = \int\limits_0^\infty \int\limits_0^\pi \frac{K^{-1}(1-\cos x) + y - \gamma^*}{t_{2D}^2} \cdot \frac{\exp\left\{\left[y + K^{-1}(1-\cos x) - \gamma^*\right]/t_{2D}\right\}}{\left\{\exp\left\{\left[y + K^{-1}(1-\cos x) - \gamma^*\right]/t_{2D}\right\} + 1\right\}^2} \times$$
$$\times \frac{y\left[y + K^{-1}(1-\cos x)\right] dxdy}{\sqrt{2y + 4\pi K^{-2} n_0 a^3 \sin^2 x}} +$$
$$+ \left\{ \int\limits_0^\infty \int\limits_0^\pi \frac{y\left[y + K^{-1}(1-\cos x) - \gamma^*\right] \exp\left\{\left[y + K^{-1}(1-\cos x) - \gamma^*\right]/t_{2D}\right\}}{\left\{\exp\left\{\left[y + K^{-1}(1-\cos x) - \gamma^*\right]/t_{2D}\right\} + 1\right\}^2 \sqrt{2y + 4\pi K^{-2} n_0 a^3 \sin^2 x}} dxdy \right\} \times$$
$$\times \left\{ \int\limits_0^\infty \int\limits_0^\pi \frac{y \exp\left\{\left[y + K^{-1}(1-\cos x) - \gamma^*\right]/t_{2D}\right\}}{\left\{\exp\left\{\left[y + K^{-1}(1-\cos x) - \gamma^*\right]/t_{2D}\right\} + 1\right\}^2 \sqrt{2y + 4\pi K^{-2} n_0 a^3 \sin^2 x}} dxdy \right\}^{-1} \times \quad (4)$$
$$\times \int\limits_0^\infty \int\limits_0^\pi \frac{K^{-1}(1-\cos x) + y - \gamma^*}{t_{2D}} \cdot \frac{\exp\left\{\left[y + K^{-1}(1-\cos x) - \gamma^*\right]/t_{2D}\right\}}{\left\{\exp\left\{\left[y + K^{-1}(1-\cos x) - \gamma^*\right]/t_{2D}\right\} + 1\right\}^2} \times$$
$$\times \frac{y\left[y + K^{-1}(1-\cos x)\right] dxdy}{\sqrt{2y + 4\pi K^{-2} n_0 a^3 \sin^2 x}}$$

In formulae (2) – (4) we introduce the following notation: $c_l$ – inversely proportional to temperature lattice thermal conductivity of SL TEM, $l$ – inversely proportional to temperature mean free path of charge carriers in SL TEM, $\zeta_{02D} = n_0 a h^2 / 4\pi m^*$, $n_0$ – bulk concentration of charge carriers in SL TEM, $K = \zeta_{02D}/\Delta$, $k_B$ – Boltzmann constant, $t_{2D} = kT/\zeta_{02D}$. Normalized to the value $\zeta_{02D}$, chemical potential $\gamma^*$ of the subsystem of free charge carriers in SL TEM is found from the equation of constancy of their concentration:

$$\frac{t_{2D}}{\pi} \int\limits_0^\pi \ln\left[1 + \exp\left(\frac{\gamma^* - K^{-1}(1-\cos x)}{t_{2D}}\right)\right] - 1 = 0. \quad (5)$$

Note that the thermal conductivity was determined on condition of the absence of current. However, in this case the Peltier heat was disregarded, as long as from our point of view we had one type of conductivity, for instance electron [17].

The results of calculations of dimensional thermoelectric figure of merit of "model" SL TEM with electron effective mass $m^* = m_0$, bulk concentration of free charge carriers $n_0 = 3 \cdot 10^{19} \text{cm}^{-3}$ with different levels of openness of FS $0.01 \leq K \leq 1$ and two different values of $a$ for temperatures $T = T_c = 300 \text{K}$, $T = T_h = 500 \text{K}$, as well as generator efficiency based on this SL TEM are given in Fig.1.

Note that the values $K < 1$ correspond to closed FS, $K = 1$ – to transient FS, and $K > 1$ – to open FS. To the so-called "two-dimensional" case correspond the values $K \gg 1$.

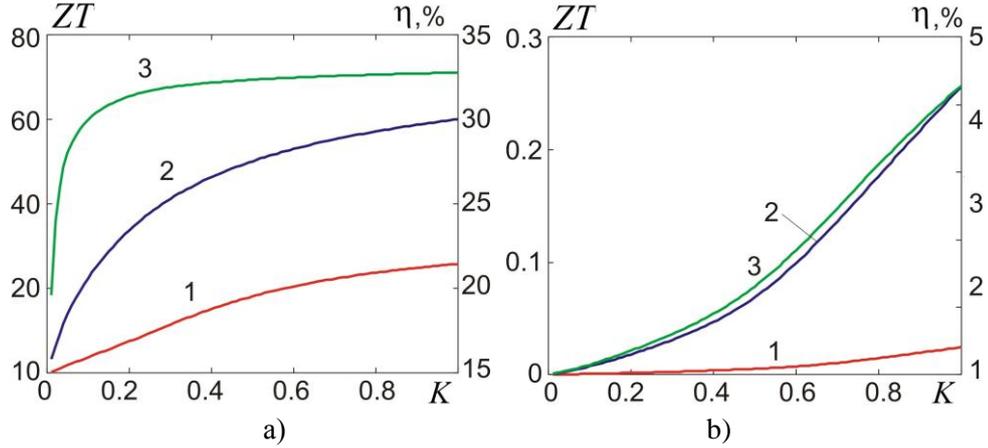

*Fig.1. Dependences of dimensionless thermoelectric figure of merit of SL TEM at $T = 300$ K (curves 1) and $T = 500$ K (curves 2), as well as generator efficiency at corresponding extreme temperatures (curves 3) with the distance between the layers of SL TEM 3nm (a) and 15nm (b), respectively.*

The generator efficiency in the construction of plots was calculated by the formula:

$$\eta = \frac{1}{1 + 2\sqrt{0.5(Z_c T_c + Z_h T_h)(Z_h T_h)^{-1}}} \cdot \frac{T_h - T_c}{T_h}. \tag{6}$$

From the figure it is seen that the figure of merit of SL TEM and the respective generator efficiency increase sufficiently fast with increasing level of openness of FS. For the considered model SL TEM the dimensionless thermoelectric figure of merit at $K = 1$ and $a = 3$ nm can reach 30-70, but already at $a = 15$ nm it drops to the values not exceeding 0.25 even at 500K. The respective generator efficiency at $a = 3$ nm and $K = 1$ reaches 34%, or 0.85 of the Carnot cycle efficiency, but at $a = 15$ nm it drops to 4%, or 0.1 of the Carnot cycle efficiency. Thus, both the level of openness of FS and the distance between the layers are important parameters defining the expedience and prospects of practical application of particular SL TEM.

Quite similarly the refrigerator coefficient of performance was found between temperatures $T_c = 230$ K, $T_h = 300$ K. In this case the following formula was used:

$$\varepsilon = \frac{\sqrt{1 + 0.5(Z_c T_c + Z_h T_h)} - T_h/T_c}{\sqrt{1 + 0.5(Z_c T_c + Z_h T_h)} + 1} \cdot \frac{T_c}{T_c - T_h}. \tag{7}$$

The results of these calculations are given in Fig.2.

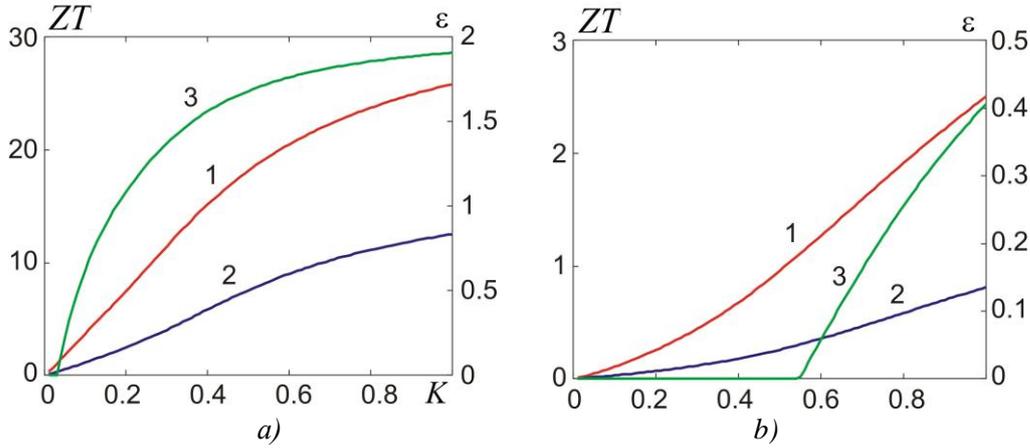

*Fig.2. Dependences of dimensional thermoelectric figure of merit of SL TEM at $T = 300$ K (curves 1) and $T = 230$ K (curves 2), as well as coefficient of performance of a refrigerator at the respective extreme temperatures (curves 3) with the distance between the layers of SL TEM 3nm (a) and 6 nm (b), respectively.*

From Fig.1 it is seen that with the optimal combination of SL TEM parameters, coefficient of performance of a refrigerator at the hot side temperature 300K and temperature difference 70K rises sharply with increase in quasi-two-dimensionality degree of SL TEM, and with the optimal combination of parameters in case of a transient FS it can reach 1.9. However, with increasing distance between the layers beyond the optimum by a factor of 2, i.e. to 6 nm, coefficient of performance for the case of a transient FS drops to 0.4. In both considered cases for given temperature difference there exists a threshold value of the level of openness of SL TEM FS below which the coefficient of performance is zero. With increasing the distance between the layers of SL TEM, this threshold value increases dramatically.

## On the reasons for low efficiency of the existing or investigated SL TEM and possible parameters of promising SL TEM

From the foregoing calculation data we see that to achieve high figure of merit of SL TEM, there must be some optimal combination of its parameters whereby the level of openness of its FS is at least close or equal to unity, and the distance between the layers is relatively small. However, the existing most popular layered and superlattice materials based on Al-Ga-As, Si-Ge or Bi(Sb)-Te(Se) systems, including those described by the Fivaz model, possess the opposite property: the distance between their layers is sufficiently large, and the level of openness of FS is low. This property is well suited for special electronics [9], including materials described by the Fivaz model, used in specific components, devices and systems, e.g. UHF-amplifiers and generators of electromagnetic waves, optical systems with self-induced transparence, soliton and breather generators, filters and delay lines, etc. However, such materials, as we can see, are not always fit for thermoelectric applications, at least as far as generators and coolers are concerned. Thus, a correct approach in this case should consist not in simple transfer of *existing* superlattice materials from electronics to thermoelectricity, but in a search for or creation of *special SL TEM,* particularly tellurium-free. This is the more relevant that the resources of tellurium-containing ores and minerals on the Earth are limited, and tellurium is just a *by-product* of their treatment, whereas the scope of application of thermoelectricity, hence production of thermoelectric modules in the world is constantly expanding. Moreover, the main consumer of tellurium today (more than 50% of available quantity) is metallurgy, rather than thermoelectricity. This raises the question of whether, at least conceptually, there can exist high-figure-of-merit SL TEM with large distances between the layers. Our analysis have shown that if free carrier concentration in the model SL TEM under study is reduced to $n_0 = 3 \cdot 10^{16} \text{cm}^{-3}$, then at $a = 30 \text{nm}$, $m^* = m_0$, $K = 1$ (it corresponds to miniband half-width $\Delta = 2.16 \cdot 10^{-4} \text{eV}$ and the ratio of longitudinal effective mass of charge carriers to transverse one, equal to 0.393) between the temperatures of 300 and 500K one can get generator efficiency 37.5%, which is very close to the Carnot cycle efficiency. As for coefficient of performance between the temperatures of 230 and 300K, it can be increased to 2.92. However, the FS of such highly efficient general-purpose SL TEM should be of the form shown in Fig.3, and its band parameters – possess high temperature stability.

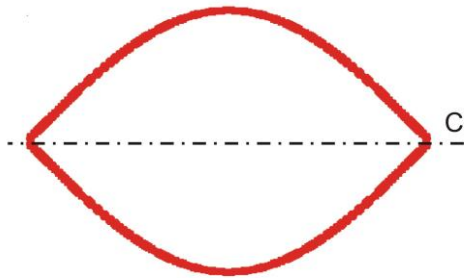

*Fig.3.The Fermi surface of possible promising high-figure-of-merit SL TEM. Cross-section of such FS by a plane parallel to SL TEM layers is a circle, the C axis is normal to the layers.*

Calculations also show that such SL TEM, if only it was created, could be also applied for the subsequent deeper cooling, for instance, from 230 to 100K, but in this case coefficient of performance would be 0.61.

In order to compare the proposed (even though hypothetical) SL TEM to the existing ones, let us note that in the overwhelming majority their miniband half-width is of the order of 0.06-0.1eV, which exceeds by a factor of 300-500 the proposed calculated value of this half-width. Moreover, for the operation of electronic devices basically using the specific nonparabolicity of band spectrum described by the Fivaz model, there is no need for complete occupancy of a narrow miniband at absolute zero temperature. The point is that for the operation of said devices it is important to have comparability of miniband half-width not to the Fermi energy, but to the energy gained by charge carrier from the electric field during superlattice period, i.e. at the distance between the neighbouring identical layers.

It can be shown that creation of high-efficiency SL TEM based, for instance, on silicon, is impossible in principle. Indeed, for this to be possible, the distance between the layers of SL TEM should satisfy the condition $a = 4\pi m^* \Delta / n_0 h^2$. In so doing, $\Delta$, at any rate, is a value of the order of forbidden energy band $E_g$, and even higher, since in traditional calculations $\Delta$ is assumed to be infinite, and FS appears to consist of equal ellipsoids. Therefore, substituting known parameters for silicon, namely $m^* = 1.26 m_0$, $E_g = 1.21 \text{eV}$, at $n_0 = 10^{15} \text{cm}^{-3}$ we obtain $a = 6.3 \text{mm}$ (!!!). It is clear that such SL TEM cannot exist. The situation will not change much even if, for instance, $n_0 = 10^{19} \text{cm}^{-3}$. In this case there must be $a = 210 \text{nm}$, while SL TEM based on Si-Ge have the highest $a$ about 30nm, and their highest dimensionless thermoelectric figure of merit does not exceed 0.75 at $a = 7 \text{nm}$ and concentration $n_0 = 3 \cdot 10^{19} \text{cm}^{-3}$, which is considered to be an order higher than the dimensionless thermoelectric figure of merit of the bulk samples. In the former case the level openness of FS is 0.16, and in the latter – 0.011. Naturally, this is very far from the required value of Q2D of high-efficiency SL TEM equal to unity.

Exactly as if at $n_0 = 3 \cdot 10^{19} \text{cm}^{-3}$, $a = 3 \text{nm}$, $m^* = m_0$ bismuth telluride was a one-valley SL TEM obeying the Fivaz model, the level of openness of its FS would be $K = 0.81\pi$ and the value of dimensionless thermoelectric figure of merit at 300K would be equal to 30.9. However, in reality, taking into account that forbidden energy band of this material is 0.13eV, and, consequently, the allowed energy band is of the order of 1.3eV (this is necessary for the validity of conventional approaches based on the parabolic band spectrum), we have $K = 0.017$, and, hence, the value of dimensionless thermoelectric figure of merit at 300K, equal to 0.531, which is quite correlated with the experimental data [18] for generator materials. For cooling materials of *Bi(Sb)-Te(Se)* system this value is equal to 0.78 [19], which is matched by $K = 0.024$. Thus, nonparabolicity in cooling materials is somewhat more pronounced than in generator materials. But these values are almost 40-60 times lower than the thermoelectric figure of merit of "genuine" high-efficiency SL TEM. It appears that conventional generator and cooling materials are also SL TEM, but with a rather low level of openness of FS, hence nonparabolicity.

Thus, we can see that the relatively low efficiency of the existing SL TEM is attributable to the fact that they do not possess a combination of parameters necessary for high figure of merit, and their FS are considerably different from that shown in Fig.3. High-efficiency SL TEM must differ from the existing in that their minibands should be so narrow that nonparabolicity described by the Fivaz model be well-expressed at low concentrations of free charge carriers. On the contrary, forbidden gaps must be sufficiently wide, in order not to "get" into intrinsic region in generation mode, since it can reduce the thermoelectric figure of merit of material. The author understands that this requirement is apparently difficult to realize technologically, but he is unaware of some fundamental physical law or principle on the basis of which this requirement might be claimed to unrealizable.

**On the methods of search for promising SL TEM with the use of quantizing magnetic fields**

In view of the foregoing, one can recommend the following four methods of search for promising SL TEM, including those described by the Fivaz model, with the use of quantizing magnetic fields:

1) investigation of de Haas-Shubnikov oscillations in quasi-classical magnetic fields at helium temperatures [20-22], including analysis of the field dependence of oscillation amplitude;
2) selection of materials according the presence of negative longitudinal magnetoresistance area and its pronounced minimum in ultra-quantum magnetic fields, if such are attainable [20,22,23];
3) selection of materials according to the degree of roundness of power factor peak in ultra-quantum magnetic fields [20,21,24,25];
4) Selection of materials according to the extent of negative longitudinal magnetoresistance at higher temperatures [20,21,26].

The above methods should be supplemented with measurement of free carrier concentration with the aid of the Hall effect and X-ray structural control of interlayer distance.

**Conclusions**

1. High-figure-of-merit SL TEM, intended for creation of thermoelectric energy converters must have FS of special form.

2. Such SL TEM must have optimal combination of parameters, in particular, free carrier concentration and the distance between the layers, and possess rather narrow conduction minibands, so that nonparaboicity described by the Fivaz model be well expressed at low charge carrier concentrations.

3. At creation or search for such SL TEM, for the identification of their FS one can employ methods based on the investigation of longitudinal electric conductivity and power factor in quantizing magnetic fields with inductions up to several T both at helium and nitrogen or room temperatures. In so doing, quantizing magnetic field, electric field and temperature gradient must be normal to SL TEM layers.

**References**

*1*. L.I.Anatychuk, L.N.Vikhor, *Functionally Graded Materials. Thermoelectricity, Vol.IV*

2. N.S.Lidorenko, V.A.Andriyako, L.D.Dudkin, E.L.Nagayev, and O.M.Narva, On the Effect of Tunneling on the Efficiency of Thermoelectric Devices, *Doklady AN SSSR* **186**, 1295 (1969).

3. A.Casian, V.Dusciak, and Iu.Coropceanu, Huge Carrier Mobilities Expected in Quasi-One-Dimensional Organic Crystals, *Phys. Rev*. B, **66,** 165404, 1-5 (2002).

4. A.I,Casian, I.I.Balmush, and V.G.Duschak, Reduction of the Lorentz Number as a New Direction of *ZT* Increase in Quasi-One-Dimensional Organic Crystals, *J.Thermoelectricity* **3**, 19 (2011).

5. L.P.Bulat, I.A.Drabkin, V.V.Karatayev, V.B.Osvensky, and D.A.Pshenai-Severin, Effect of Interface Scattering on the Thermal Conductivity of $Bi_xSb_{2-x}Te_3$ Nanostructured Semiconductor Material, *Physics of the Solid State* **52**, 1712 (2010).

6. L.I.Anatychuk, *Thermoelectricity, Vol.1, Physics of Thermoelectricity* (Chernivtsi: Institute of Thermoelectricity, 2008), 404p.

7. J.P.Heremans, Low Dimensional Thermoelectricity. Proceedings of XXXIV International School of Semiconducting Compounds. Jaszowiec-2005, *Acta Physica Polonica* **108**(4), 609-634 (2005).

8. M.S.Dresselhaus, G.Dresselhaus, X.Sun, Z.Zhang, S.B.Cronin, and T.Koga, Low Dimensional Thermoelectric Materials, *Physics of the Solid State* **41**(5), 755-758 (1999).

9. F.G.Bass, A.A.Bilgakov, and A.P.Tetervov, *High-Frequency Properties of Superlattice Semiconductors* (Moscow: Nauka, 1989), 288p.

10. W.L.Liu, T.Borca-Tasciuk, J.L.Liu, K.Taka, K.L.Wang, M.S.Dresselhau, and G.Chen, In-Plane Thermoelectric Properties of Si/Ge Superlattice, *Proc. of 20[th] International Conference on Thermoelectrics (2001). – P.340-343.*

11. R.Venkatasubramanian, E.Siivola, and T.S.Colpiits, In-Plane Thermoelectric Properties of Freestanding *Si/Ge* Superlattice Structures, *Proc. of 17[th] International Conference on Thermoelectrics (1998). – P.191-197.*

12. A.Lambrecht, H.Beyer, J.Nurnus, C.Künzel, and H.Böttner, High Figure of Merit ZT in *PbTe* and $Bi_2Te_3$ Based Superlattice Structures by Thermal Conductivity Reduction, *Proc. of 20[th] International Conference on Thermoelectrics* (2001). – P.335-339.

13. B. Yang, J. Liu, K. Wang, and G. Chen, Characterization of Cross-Plane Thermoelectric Properties of *Si/Ge* Superlattices, *Proc. of 20[th] International Conference on Thermoelectrics (2001),* 344-347.

14. Y.Zhang, D.Vashaee, R.Singh, and A.Shakouri, Influence of Doping Concentration and Ambient Temperature on Cross-Plane Seebeck Coefficient of InGaAS/InAlAs Superlattices, *Mat. Res. Soc. Symp. Proc.* **793**, 59-65 (2004).
15. J.C.Caylor, K.Coolney, J.Stuart, S.Nangoy, T.Colpiits, and R.Venkatasubramanian, Developing PbTe-Based Superlattice Structures with Enhanced Thermoelectric Performance, *2005 International Conference on Thermoelectrics*, P.489-491.
16. R.F.Fivaz, Theory of Layered Structures, *J. Phys. Chem. Solids* **26**(5), 839-845 (1967).
17. N.F.Hinsche, I.Mertig, and P.Zahn, Lorentz function of $Bi_2Te_3/Sb_2Te_3$ Superlattices, *J. Electron. Mat.* **42**(7), 1406-1410 (2013).
18. B.M.Goltsman, V.A.Kudinov, and I.A.Smirnov, *Semiconductor Thermoelectric Materials Based on $Bi_2Te_3$* (Moscow: Nauka, 1972), 320p.
19. L.D.Ivanova, Yu.V.Granatkina, A.Dauscher, B.Lenoir, and H.Sherrer, Influence of the Purity and Perfection of Czochralski-Grown Single Crystals of Bismuth and Antimony Chalcogenides Solid Solution on Their Thermoelectric Properties, *Proc. of $5^{th}$ European Workshop on Thermoelectrics, Pardubice, Czech Republic, 1999*, P.175-178.
20. P.V.Gorskyi, (New York: *Layered Structure Effects as Realization of Anisotropy in Magnetic, Galvanomagnetic and Thermoelectric Phenomena* Nova Publishers, 2014), 366p.
21. P.V.Gorskyi, Diagnostics of Functional Materials with Closed Fermi Surfaces Described by the Fivaz Model and Some Aspects of Their Use, *Dopovidi NAN Ukrainy* 12, 77-85 (2014).
22. P.V.Gorskyi, Electric Conductivity of Functional, Including Thermoelectric, Materials Described by the Fivaz Model, in Quasi-Classical Range of Magnetic Fields, *J.Thermoelectricity* **3**, 5-14 (2014).
23. P.V.Gorskyi, Are Layered Structure Effects Manifested with Close Fermi Surfaces? *Ukrainian J. Physics* **55**(12), 1297-1305 (2010).
24. P.V.Gorskyi, Power Factor of Layered Thermoelectric Material with Closed Fermi Surface in a Quantizing Magnetic Field, *Ukrainian J. Physics* **58**(4),1297-1305 (2013).
25. P.V.Gorskyi, Diagnostics of Functional Materials Described by Fivaz Model by the Dependence of Power Factor on Quantizing Magnetic Field. – *Collected Papers Based on Proc. of XX International Remote Scientific and Practical Conference "Scientific Discussion. Problems of Mathematics, Physics, Chemistry, Biology", №8(19), Moscow-2014* (Moscow: Nauka, Interperiodica), P..55-60.
26. P.V.Gorskyi, *Gigantic Negative Magnetoresistance of Nanoheterostructures, Described by Fivaz Model.* – arXiv:1503.06823v1 -10Mar 2015. – P.1-3.

The Author is grateful to academician L.I.Anatychuk for the formulation of the problem, as well as to L.N.Vikhor, chief research scientist, for the helpful discussion of the results of the work.